\documentclass[12pt]{SciPost}

\hypersetup{
    colorlinks,
    linkcolor={red!50!black},
    citecolor={blue!50!black},
    urlcolor={blue!80!black}
}

\usepackage[bitstream-charter]{mathdesign}
\DeclareSymbolFont{usualmathcal}{OMS}{cmsy}{m}{n}
\DeclareSymbolFontAlphabet{\mathcal}{usualmathcal}

\fancypagestyle{SPstyle}{
\fancyhf{}
\lhead{\colorbox{scipostblue}{\bf \color{white} ~SciPost Physics Core }}
\rhead{{\bf \color{scipostdeepblue} ~Submission }}

\fancyfoot[C]{\textbf{\thepage}}
}

\usepackage{amsmath}
\usepackage{graphics}
\usepackage{mymacros}
\usepackage{braket}
\usepackage{ytableau}

\usepackage{bm}

\newcommand{\pochb}[2]{(#1)^{\overline{#2}}}
\newcommand{\poch}[2]{#1^{\overline{#2}}}
\renewcommand{\l}{\lambda}

\newcommand{\supp}{\operatorname{supp}}
\renewcommand{\h}{\operatorname{h}}

\newcommand{\colA}{*(red!25)}
\newcommand{\colB}{*(blue!25)}
\newcommand{\colC}{*(orange!25)}
\newcommand{\colD}{*(gray!30)}

\newcommand{\HGF}[3]{\, {}_3F_2\left[\begin{array}{c}#1\\#2\end{array};#3\right]}

\begin{document}

\pagestyle{SPstyle}

\begin{center}{\Large \textbf{\color{scipostdeepblue}{
Revisiting the Page curve and its moments. A combinatorial approach\\
}}}\end{center}

\begin{center}\textbf{
Gero von Gersdorff\textsuperscript{1$\star$},
}\end{center}

\begin{center}
{\bf 1} Pontifícia Universidade Católica do Rio de Janeiro,
Rua Marquês de São Vicente 225, Rio de Janeiro, Brazil
\\[\baselineskip]
$\star$ \href{mailto:email1}{\small gersdorff@puc-rio.br}
\end{center}

\section*{\color{scipostdeepblue}{Abstract}}
\textbf{\boldmath{%
We revisit the calculation of the von Neumann (or "entanglement") entropy of a subsystem of a pure quantum state, under the assumption that
the latter is drawn at random from a uniform distribution on the full Hilbert space. We derive simple and closed expressions for all power moments, from which the moments of the entropy can be computed by simple
differentiation.
Our approach (different from the usual one based on random matrix theory and Laguerre polynomials) makes use of Schur-Weyl duality and the
character theory of the symmetric group $S_N$. The paper is self-contained, providing all the necessary mathematical background.
}}

\vspace{10pt}
\noindent\rule{\textwidth}{1pt}
\tableofcontents
\noindent\rule{\textwidth}{1pt}
\vspace{10pt}

\section{Introduction}

Given a bipartite pure quantum state, it is natural to ask what is the entropy of the mixed state obtained by tracing over one of the factors. Choosing this pure state at random from some 
given distribution turns this question into a statistical one where the entropy itself becomes a random variable. What is the average entropy? Is the average entropy typical for such a state?

If the traced system B is  much larger than the untraced system A, 
 it is reasonable to expect that the 
resulting state $\rho_A$ is maximally mixed. Denoting by $m$ and $n$ the dimensions of systems A and B respectively, one thus expects the  entropy
 $S_A\sim \log(m)$ for $m\ll n$. Similarly, in the opposite limit,  the state $\rho_A$ is  close to pure, and 
 $S_A\to 0$ as $n\to 1$. In fact, using that $S_{\rm A}=S_\text{B}$, one expects $S_\text{A}\sim \min(\log(m),\log(n))$, with maybe some smooth  correction near $n\sim m$. This is the Page curve \cite{Lubkin:1978nch,Page:1993df}.

Famously, the Page curve has been conjectured to govern the entropy of black hole evaporation \cite{Page:1993wv}. Under the assumption that black hole evolution is in fact unitary, and that the black hole is formed from a pure state, the total system of Hawking radiation (A) and black hole (B) will remain pure, even though an observer measuring only the  radiation  will initially see some nonzero entropy.
As the black hole  evaporates, the entropy $S_{\text A}$ will grow, but only until  the dimensions of the subspaces A and B are of the same order (the "Page time"). After this moment, the entropy $S_{\text{A}}$ will 
fall again, until it reaches zero when the black hole has completely radiated away.
Recent progress in black hole physics supports this conjecture \cite{Penington:2019npb,Almheiri:2019psf,Almheiri:2019hni,Penington:2019kki,Almheiri:2019qdq}. 
The entanglement viewpoint is also central to modern quantum statistical mechanics:
canonical ensembles do not require a microcanonical-ensemble origin. Instead, any {\em pure} fixed-energy state, when tracing out a sufficiently large subsystem already 
leads to a thermal density matrix with overwhelming probability \cite{Goldstein:2005aib,Popescu:2006rhr}.
Therefore, thermality arises purely from entanglement. The fact that a subsystem density matrix of a random pure state is already close to its averaged value often goes by the name of {\em typicality}.
It has important consequences in quantum information theory \cite{Hayden:2005sqo}.

Page quantified the above qualitative behaviour (assuming the original  state is taken from the uniform measure on the normalized pure states), and  conjectured the general form of the average entropy to be \cite{Page:1993df}
\be
\overline {S_{\text{A}}}=H_{mn}-H_{n}-\frac{m-1}{2n}
\label{eq:page}
\ee
where w.l.o.g. $m\leq n$, and $H_k=\sum_{j=1}^k \frac{1}{j}$ are harmonic numbers.
For sufficiently large $n$ eq.~(\ref{eq:page}) converges indeed to $\log(m)$.
Page obtained eq.~(\ref{eq:page}) by evaluating the integrals over the explicit form of the measure \cite{Lloyd:1988cn} exactly in a few tractable cases and (correctly) guessed 
the general form. 
Soon after, eq.~(\ref{eq:page}) was proven rigorously, and later reinterpreted as averages of random matrix ensembles
 \cite{Foong:1994eja,Sanchez-Ruiz:1995bhf,Sen:1996ph,Sommers:2004xri}. 
The next logical question is if the mean entropy is typical. This requires the   variance  which was calculated in \cite{Vivo:2016ref,Wei:2017jyl} using  similar techniques. 
Some higher moments were computed in \cite{Wei_2020,Huang:2021btj}.
Later work \cite{Bianchi:2019stn} approached the problem via the calculation of the 
power moments of $\rho_{\text A}$, which allows one to obtain the moments of $S_{\text A}$ by differentiation (see also ref.~\cite{Sommers:2004xri}). The same authors also gave an interesting 
extension to bipartite pure states that are not direct products, for instance  states of fixed energy.

In this work we propose a purely algebraic approach to the calculation of the moments of the subsystem entropy. 
Our main focus are power moments of the subsystem density matrix. We will use $U(mn)$ invariance of the
uniform measure to compute these quantities without ever evaluating any integrals. 
The problem becomes a purely group theoretical/combinatorial one, and is a nice application of various concepts in these fields, in particular Schur-Weyl duality and the Murnaghan–Nakayama formula (all necessary mathematical background will be provided in a self-contained way). 
The drawback of the method is that it a priori only gives integer power moments, and the continuation to 
real powers is not completely mathematically rigorous. However, consistency checks and monotonicity constraints basically fix the continuation. We also check our results against the literature, finding agreement, while providing considerably simpler expressions. 

A similar approach has been advocated in \cite{Collins:2009hos,Liu:2017lem}, however, an explicit evaluation of the resulting sums over permutations was not performed in these works.

\section{General considerations}


Consider a generic bipartite pure state 
\be
\ket {\psi}=\psi^{ia}\ket {i}_{\!\text A}\ket {a}_{\! \text B}
\ee
where $i$ and $a$ label orthonormal bases for the two systems A and B. Let their dimensions be $m$ and   $n$ respectively, the total dimension will be denoted by $d=mn$.
After tracing over B,  system A is in a mixed state, described by the density matrix 
\be
\rho_{\text A,\psi}=\psi \psi^\dagger
\ee
Its von Neumann entropy is given by
\be
S_{\text A,\psi}=-\tr\, \rho_{\text A,\psi}\,\log\rho_{\text A,\psi}
\ee

Let $\psi$ be distributed by the uniform measure on the total Hilbert space, i.e.,  the uniform measure on the manifold $\mathbb{CP}^{d-1}$ of normalized states,
$d\mu_\psi\sim d^d\psi d^d \psi^*\delta(1-\tr[\psi\psi^\dagger])$. This  induces a measure on the eigenvalues of the matrix $\psi\psi^\dagger$ \cite{Lloyd:1988cn}, which is usually employed to evaluate the expectation value and higher moments of $S_{\text A,\psi}$ \cite{Page:1993df,Foong:1994eja,Sanchez-Ruiz:1995bhf,Sen:1996ph,Sommers:2004xri,Vivo:2016ref,Wei:2017jyl,Wei_2020,Huang:2021btj,Bianchi:2019stn}.
The only property of the measure that matters 
for the rest of this paper is its invariance under U$(d)$ rotations, in particular, its explicit form is irrelevant and we will never actually perform any integration.

The average entanglement entropy of the system A with respect to this distribution is
\be
\overline{S_{\text  A}}\equiv \int d \mu_\psi\, S_{\text A,\psi}
\ee
and its higher moments are defined accordingly.
A standard trick to calculate the  entropy exploits the identity 
\be
\tr \rho \log \rho=\frac{d}{d\l} \tr \rho^\l  \big|_{\l=1}
\label{eq:master}
\ee
We will refer to the averages of the type $\overline{\tr\rho^\l_{\text{A}}}$ as  power moments.
In general, the $k^\text{th}$ moment of the entropy can be obtained by differentiation of the 
generalized power moments
\be
\overline{\tr [\rho_{\text A }^{\lambda_1}]\cdots \tr [\rho_{ \text A }^{\lambda_k}]}=
\int d\mu_\psi\, \tr [\rho_{\text A,\psi }^{\lambda_1}]\cdots \tr [\rho_{\text  A,\psi }^{\lambda_k}]
\label{eq:trrhoN}
\ee
with respect to the variables $\l_i$ at the special point $\l_i=1$.
Since the eigenvalues of $\rho_{A,\psi }$ are nonnegative real numbers, eq.~(\ref{eq:trrhoN}) is  well defined  for any real positive $\lambda_i$. In contrast to the usual approach, here we will  compute the power moments combinatorially, in particular only for arbitrary positive integers, and the final result is  to some extent ambiguous (i.e., there may be several ways of  continuing the result from the positive integers to the real numbers, and a priori it is unclear which one corresponds to the one implied by 
eq.~(\ref{eq:trrhoN})). One can however perform some nontrivial cross checks on the final result, and also compare with known exact results in the literature.
\footnote{Related issues arise in computations of the black hole entropy via the replica trick which also start from an inherently  discrete variable (the number of replicas). See e.g.~refs.~\cite{Lewkowycz:2013nqa,Dong:2016fnf,Penington:2019kki,Almheiri:2019qdq}.}

We now describe how this can be done. Using U$(d)$ invariance and obvious permutation symmetry,  we can  write 
 (temporarily switching to a composite index notation $I=\{i,a\}$) 
\be
\int d\mu_\psi\ \psi^{I_1}\psi^*_{J_1}\cdots \psi^{I_N}\psi^*_{J_N}= \frac{1}{Z_{(1^N)}}
\left(\delta^{I_1}_{J_1}\cdots \delta^{I_N}_{J_N}+ \text{permutations}\right)
\label{eq:invariance}
\ee
where the sum is over all permutations of the $N$ indices $J_k$, and $Z_{(1^N)}$ is some normalization factor (the notation will become clear in a moment).
For convenience, let us give the  tensors appearing on the rhs of eq.~(\ref{eq:invariance}) a name: \footnote{$T_d(p)$ can be naturally thought of as an endomorphism acting on the direct product vector  space $\mathbb C^{d^N}=\mathbb C^d\otimes\mathbb C^d\otimes\cdots\otimes \mathbb C^d$  by permuting the $N$ factors with $p$. Eq.~(\ref{eq:cyclelength}) is then the trace of this endomorphism over $\mathbb C^{d^N}$.} 
\be
[\,T_d(p)\,]^{I_1\dots I_N}_{J_1\dots J_N}\equiv \delta^{I_1}_{J_{p(1)}}\cdots \delta^{I_N}_{J_{p(N)}}\qquad p\in S_N
\ee
One can compute $Z_{(1^N)}$ by 
 tracing over the indices, $I_k=J_k$, using the normalization condition $\psi^I\psi^*_I=1$. 
Then the term in the bracket in eq.~(\ref{eq:invariance}) corresponding to $p$ gives
\be
\tr T_d(p) = d^{\gamma(p)}
\label{eq:cyclelength}
\ee
where  $\gamma(p)$ counts the number of cycles in the permutation $p$.
The sum over $p$ can then be evaluated immediately
\be
Z_{(1^N)}
=\sum_{p\in S_N} d^{\gamma(p)} 
=\sum_{\ell=1}^N c(N,\ell) d^\ell=\poch{d}{N}
\label{eq:normf}
\ee
Here, $c(N,\ell)$ are the Stirling numbers of the first kind, which count  the number of $p\in S_N$ with precisely $\ell$ cycles, and 
 $\poch{d}{N}$ denotes the Pochhammer symbol (or rising factorial)
\be
\poch{d}{N}\equiv 
d(d+1)\cdots (d+N-1)=\frac{\Gamma(d+N)}{\Gamma(d)}\,.
\ee
The last equality of eq.~(\ref{eq:normf}) is a standard identity for the Stirling numbers.

Next we would like to compute (for integer $N$)
\be
\overline{\tr[\rho_{\text A}^N]}= \frac{Z_{(N)}}{ Z_{(1^N)}}
\label{eq:singletrace}
\ee  
where $Z_{(N)}$ is defined by tracing the invariance condition eq.~(\ref{eq:invariance})  in a shifted way
\be
Z_{(N)}\equiv
\sum_{p\in S_N}\tr T_n(p)\,\tr T_m(pq)
\ee
Here $q$ denotes a fixed permutation that has exactly one $N$-cycle,  any
 such  permutation will produce precisely the trace on the left hand side of eq.~(\ref{eq:singletrace}).
Similarly,  
\be
\overline{\tr[\rho_{\text A}^P]\tr[\rho_{\text A}^Q]}= \frac{Z_{(P,Q)}}{ Z_{(1^N)}}
\label{eq:doubletrace}
\ee  
 where
\be
Z_{(P,Q)}\equiv 
\sum_{p\in S_N}\tr T_n(p)\,\tr T_m(pq')
\ee
where $q'$ denotes any permutation with one $P$ and one $Q$ cycle and we assume $P+Q=N$.
 
Note that  $Z_{(N)}$ and $Z_{(P,Q)}$ are class functions of $q$ and $q'$ respectively, i.e., they only depend on 
the cycle types of $q$ and $q'$, which in turn precisely correspond to  conjugacy classes of $S_N$.
They  are labeled by  partitions of $N$, i.e., $(N)$ and  $(P,Q)$ respectively.
In the following, conjugacy classes/partitions of $S_N$  are denoted by lowercase Greek letters.
Note that they simultaneously also label a Young diagram (see appendix \ref{sec:SN} for a quick summary of the group $S_N$).

It is now obvious how to continue this for higher moments. The integral in eq.~(\ref{eq:trrhoN}) is given by $Z_{\lambda}/Z_{{(1^N)}}$, where
\be
{Z_\lambda}=\sum_{p\in S_N} \tr T_n(p)\, \tr T_m(pq)
\qquad q\in C_\lambda
\label{eq:Zlambda}
\ee
Here $\lambda=(\lambda_1,\lambda_2,\dots \lambda_k)$ is a partition of $N$ with $\lambda_i\geq\lambda_{i+1}$, and $q$ is any element of the associated conjugacy class $C_\l$, i.e., a permutation of cycle type $\lambda$. Notice that ${(1^N)}$ corresponds to the conjugacy class of the identity permutation.

Using eq.~(\ref{eq:cyclelength}) it is possible to express $Z_\l$ in terms of cycle lengths \cite{Collins:2009hos,Liu:2017lem}:
\be
{Z_\lambda}
=\sum_{p\in S_N} n^{\gamma(p)}m^{\gamma(pq)}\qquad q\in C_\lambda
\label{eq:Zlambda2}
\ee
This  can easily be implemented 
on a computer but  is still very non-explicit. Moreover,  it remains unclear how to continue it to 
real $\lambda_i$. We will instead use the expression  eq.~(\ref{eq:Zlambda}) as a starting point in the next section.


\section{Derivation of the main result}

For arbitrary $\lambda$, eq.~(\ref{eq:Zlambda2}), cannot be evaluated as straightforwardly as in  eq.~(\ref{eq:normf}).
Instead, we can rewrite eq.~(\ref{eq:Zlambda}) directly,  making  use of Schur-Weyl duality. We find (see appendix \ref{sec:SW})
\be
Z_\lambda=\sum_{\mu\,\vdash N} \chi_\mu(\lambda)\, \frac{C_\mu(n)C_\mu(m)}{H_\mu}
\label{eq:level2}
\ee
where the sum is over all Young diagrams $\mu$ of size $N$ with at most $\min(m,n)$ rows, and
$\chi_\mu(\lambda)$ is the character of the partition $\lambda$ in the representation labeled by $\mu$.
 Furthermore, we defined the content polynomials 
\be
C_\mu(n)\equiv\prod_{(i,j)\in \mu}(n+j-i)
\label{eq:content}
\ee
and  the product of hook lengths
\be
H_\mu\equiv\prod_{(i,j)\in \mu}h_{i,j}
\label{eq:hook}
\ee
where $h_{i,j}$ is the hook length of box $(i,j)$ of the Young diagram,  see  appendix \ref{sec:SN}. 
In the rest of this section we will perform the sum over $\mu$ which will turn eq.~(\ref{eq:level2}) into a completely explicit expression.

\begin{figure}
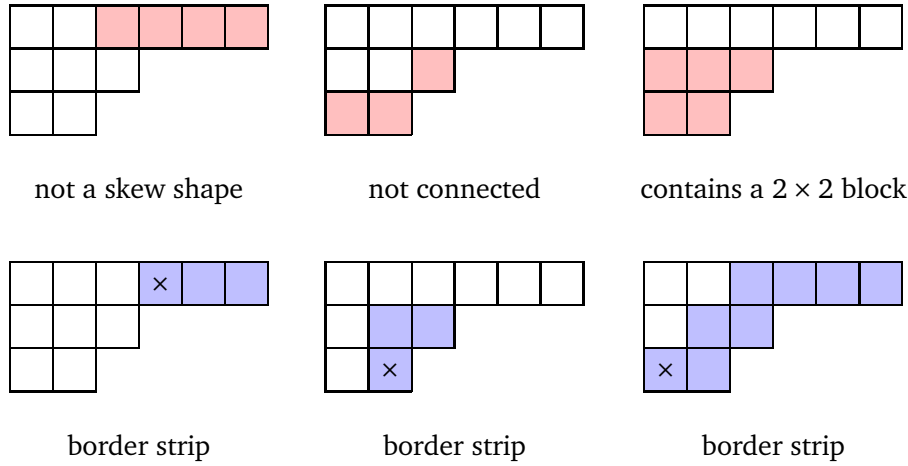

\[
\renewcommand{\arraystretch}{2.5}
\begin{array}{c@{\qquad}c@{\qquad}c}

\begin{ytableau}
{} & {} & \colA & \colA & \colA & \colA
\\
{} &  & 
\\
{} & {}
\end{ytableau}
&
\begin{ytableau}
{} & {} & {} & {} & {} & {}
\\
{} & {} & \colA{}
\\
\colA{} & \colA{}
\end{ytableau}
&
\begin{ytableau}
{} & {} & {} & {} & {} & {}
\\
\colA{} & \colA{} & \colA{}
\\
\colA{} & \colA{}
\end{ytableau}
\\[20pt]
\text{not a skew shape}
&
\text{not connected}
&
\text{contains a }2\times 2\text{ block}
\\[6pt]

\begin{ytableau}
{} & {} & {} & \colB{\times} & \colB{} & \colB{}
\\
{} & {} & {}
\\
{} & {}
\end{ytableau}
&
\begin{ytableau}
{} & {} & {} & {} & {} & {}
\\
{} & \colB{} & \colB{}
\\
{} & \colB{\times}
\end{ytableau}
&
\begin{ytableau}
{} & {} &\colB {} & \colB{} & \colB{} &\colB {}
\\
{} & \colB{} & \colB{}
\\
\colB{\times} & \colB{}
\end{ytableau}
\\
\text{border strip}
&
\text{border strip}
&
\text{border strip}
\end{array}
\]
\caption{Some subsets of the Young diagram of $(6,3,2)$. Sets in red  violate some of the 3 conditions for border strips, while sets in blue  satisfy all of them. Lower left corners (LLCs) are marked with a cross.}
\label{fig:boarderstrips}
\end{figure}

The characters $\chi_\mu(\lambda)$, defined as the trace of (any member of the conjugacy class labeled by) $\lambda$ in 
the representation $\mu$, can be computed by the Murnaghan–Nakayama (MN) rule \cite{Murnaghan:1937,Nakayama:1940a,Nakayama:1940b,macdonald1998symmetric,stanley1999enumerative}.
The essential ingredient of this method are so-called border strips which we now quickly review. 
For any Young diagram $\mu$, a border strip is defined to be a set of boxes of $\mu$ satisfying  the following conditions
\begin{enumerate}
\item
Removing the boxes of the border strip leaves a valid Young diagram
\item
The set is connected
\item
The set contains no $2\times 2$ block
\end{enumerate}
Sets only satisfying the first condition are called skew-shapes.
The name border strip arises from the fact that the set necessarily lies at the outer (right) border of the
diagram $\mu$. The length of a border strip is the number of boxes it occupies, 
its height  is defined to be the number of rows it occupies minus one.
Examples of border strips are shown in figure \ref{fig:boarderstrips}.

\begin{figure}
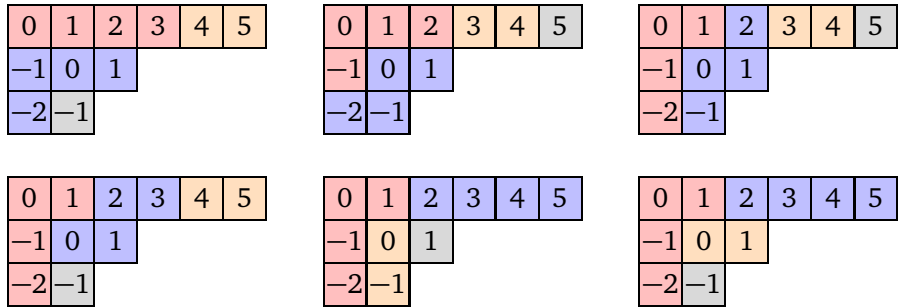


\[
\renewcommand{\arraystretch}{2.7}
\begin{array}{c@{\qquad}c@{\qquad}c}

\begin{ytableau}
\colA0 & \colA1 & \colA2 & \colA3 & \colC4 & \colC5\\
\colB-1 & \colB0 & \colB1\\
\colB-2 & \colD -1
\end{ytableau}
&
\begin{ytableau}
\colA0 & \colA1 & \colA2 & \colC3 & \colC4 & \colD 5\\
\colA-1 & \colB0 & \colB1\\
\colB-2 & \colB-1
\end{ytableau}
&
\begin{ytableau}
\colA0 & \colA1 & \colB2 & \colC3 & \colC4 & \colD 5\\
\colA-1 & \colB0 & \colB1\\
\colA-2 & \colB-1
\end{ytableau}
\\[28pt]
\begin{ytableau}
\colA0 & \colA1 & \colB2 & \colB3 & \colC4 & \colC5\\
\colA-1 & \colB0 & \colB1\\
\colA-2 & \colD -1
\end{ytableau}
&
\begin{ytableau}
\colA0 & \colA1 & \colB2 & \colB3 & \colB4 & \colB5\\
\colA-1 & \colC0 & \colD 1\\
\colA-2 & \colC-1
\end{ytableau}
&
\begin{ytableau}
\colA0 & \colA1 & \colB2 & \colB3 & \colB4 & \colB5\\
\colA-1 & \colC0 & \colC1\\
\colA-2 & \colD -1
\end{ytableau}
\end{array}
\]
\caption{The border strip decompositions of the partition $\mu=(6,3,2)$ into border strips of lengths 
$\lambda=(4,4,2,1)$. The numbers in the cells correspond to their content, defined as column number minus row number. Evaluating the heights of these strips, they are seen to contribute $-1$, $+1$, $+1$, $-1$, $-1$, $+1$ and hence the character $\chi_\mu(\lambda)$ is zero in this example.}
\label{fig:MNrule}
\end{figure}

The MN rule now states the following. In order to compute the character $\chi_\mu(\lambda)$, remove first a border strip of length $\lambda_k$ from $\mu$ to produce a smaller Young diagram, then remove a border strip from that diagram of length $\lambda_{k-1}$ and so on, until only the empty diagram is left. This way, $\mu$ is  written as a union of border strips. 
Equivalently, successively add border strips, starting from the empty partition, in order to arrive at the diagram $\mu$ (this reverse procedure is  more useful  for our purposes).
The decomposition is not unique, in fact there may exist more than one or none at all. The character is simply computed as 
\be
\chi_\mu(\lambda)=\sum_{\text{decomp.}}(-1)^{\h (s_1)+\dots+\h(s_k)}
\ee
where the sum is over all border strip decompositions of $\mu$ of type $\lambda$,  $s_i$ are the strips of each decomposition, and $\h (s_i)$ denotes the height of the border strip. An example is provided in figure \ref{fig:MNrule}.

We are now in the position to evaluate the character sums eq.~(\ref{eq:level2}). 
The simplest ones turn out to be the ones with a single cycle $\lambda=(N)$, which is
the relevant quantity for the calculation of the mean of the entropy, i.e., the Page curve.
We will then proceed to evaluate the case $\lambda=(P,Q)$, and finally derive the expression for the  general case.

\subsection{$\lambda=(N)$}

Only the $\mu$'s corresponding to hook partitions contribute in the MN formula.
We can label them as $\rho{(r)}=(N-r,1^r)$. These $\rho(r)$ are already border strips  of length $N$  and height $r$, and therefore their
 character is  
 \be
 \chi_{\rho(r)}(\lambda)=(-1)^r\,.
 \ee
Moreover, the hook product has three factors
\be
H_{\rho(r)}=N(N-r-1)!r!
\ee
corresponding to the cell  (1,1), the cells to its right, and the cells below it respectively. 
Finally, the content polynomial is
\be
C_{\rho(r)}(n)=(n-r)^{\overline{N}}\,.
\ee
Therefore, 
defining
\be
F_1(r,N)\equiv 
(-1)^r\frac{(n-r)^{\overline{N}}
(m-r)^{\overline{N}}}{N(N-r-1)!r!}
\label{eq:defF}
\ee
we can write the result immediately as
\be
Z_{(N)}
=\sum_{r\geq 0}F_1(r,N)
\ee
where we have dropped the upper limit as $F_1(r,N)=0$ when $r\geq N$ because of the hook product. Notice that the sum also terminates at $\min(m,n)-1$ because of the content polynomials, a property that will remain true when we take $N$ real.

\subsection{$\lambda=(P,Q)$}

This case is relevant for the variance of the Page curve. 
To compute $\chi_\mu(\lambda)$, we must decompose $\mu$ into border strips of lengths $P$ and $Q$.

In order to evaluate the sum in eq.~(\ref{eq:level2}), we need a good coordinate system for the Young diagrams $\mu$. In the case $\lambda=(N)$ only the hook-shape partitions gave nonzero characters, and it was possible to parametrize them by their height $r$, where $0\leq r< N$, this defined a map $\rho(r)$. Notice that $-r$ corresponded to the content of the cell in the lower left corner (LLC). In particular 
$C_{\rho(r)}(n)=(n-r)^{\overline N}$ was a simple rising factorial.
It turns out that the "good" generalization to the case $\lambda=(P,Q)$ is to take as coordinates the contents of the LLCs of the two  border strips contributing in the MN rule. 

We thus define a map $\rho(r,s)$ from $D_\rho\subset\mathbb  Z^2$ onto the set of Young diagrams as follows.
Consider all border strips of length $P$, with LLC cell content $-r$, where $0\leq r< P$, and to each add all border strips of length $Q$ with LLC content $-s$. 
The possible allowed values of $s$ depend on $r$, a useful way of bookkeeping is via so-called beta sets (see app.~\ref{sec:beta}, where we also present some examples of this construction).
For a given $s$ there exists at most one such strip, hence the map $\rho$ is well defined.
The explicit form of $D_\rho$ is given in eq.~(\ref{eq:Drho2}).
\footnote{
We mention in passing that the order of adding strips does not matter (as long as the same order is used consistently),  i.e., we could start by adding a border strip of length $Q$ first and after that a border strip of length $P$, this would change both the map $\rho$ as well as the domain $D_\rho$, but the result for the characters (and hence the sum) is the same by the MN rule.}

The map $\rho$ allows us to change the sum over Young diagrams to a sum over $(r,s)\in D_\rho$ 
\be
\begin{aligned}
Z_{(P,Q)}={}&\sum_{\mu} \chi_\mu(\lambda)\ \frac{C_\mu(m)C_\mu(n)}{H_\mu}\\
={}&
\sum_{\mu} \biggl(\sum_{(r,s)\in \rho^{-1}(\mu)} (-1)^{\h (r)+\h (s)}\biggr)\  \frac{C_\mu(m)C_\mu(n)}{H_\mu}\\
={}&\sum_{(r,s)\in D_\rho}(-1)^{\h (r)+\h (s)}\ \frac{C_{\rho(r,s)}(m)C_{\rho(r,s)}(n)}{H_{\rho(r,s)}}
\label{eq:rewriting}
\end{aligned}
\ee
where in the second equality we used the MN formula to compute $\chi_{\mu}(\lambda)$.


The $(r,s)$ coordinates have the nice feature that
\be
C_{\rho(r,s)}(n)=(n-r)^{\overline{P}}(n-s)^{\overline{Q}}
\ee
by the definition of the $r$ and $s$ variables as the content of the inner corner of the strip, and the simple fact that border strips have "continuous" content $-r,-r+1,\dots, -r+P-1$ etc, which follows directly from the definition of the border strip, see also figure \ref{fig:MNrule}.
Moreover, by explicitly evaluating the hook product using beta-sets, we find the closed expression eq.~(\ref{eq:hookprodsigned}) and hence
\be
Z_{(P,Q)}=\sum_{(r,s)\in D_\rho}F_1(r,P)F_1(s,Q)X(r,s,P,Q)\equiv 
	\sum_{(r,s)\in D_\rho}F_2(r,s;P,Q)
\ee
where $F_1$ was defined in eq.~(\ref{eq:defF}) and $X$ is the cross ratio
\be
X(r,s,P,Q)\equiv \frac{
(s-r)(Q-P-s+r)
}{
(P-r+s)(Q-s+r)
}
\label{eq:defX}
\ee

The set $D_\rho$ is still not very convenient to work with in practice. 
However, eq.~(\ref{eq:rel1}) implies that
\be
\begin{aligned}
\sum_{\substack{0\leq r<P\\s=r-P}} F_2(r,s;P,Q)={}&
\sum_{\substack{0\leq s<Q\\ r=s+P}}F_2(r,s;P,Q) 
+\sum_{0\leq u<P+Q} F_1(u,P+Q)
\end{aligned}
\label{eq:posneg}
\ee
Using this to eliminate the negative values of $s$ in favor of positive values of $r$, 
we obtain a new domain which equals 
$\supp(F_{2})\cap (\mathbb Z_{\geq0})^2$,
plus a new one-dimensional domain which equals $\supp(F_{1})\cap \mathbb Z_{\geq0}$ (from the second term in eq.~(\ref{eq:posneg})). 
\footnote{To be more precise, define $G_{2}(r,s)\equiv 1/H_{\rho(r,s)}$, and $G_1(u)=1/H_{\rho(u)}$. The statement about the support is actually true for $G_2$ and not $F_2$: for $\min(m,n)\geq P+Q$ the supports match $\supp(G_2)=\supp(F_2)$, otherwise $\supp(G_2)\supset\supp(F_2)$. This does however not affect the result eq.~(\ref{eq:finalPQ}). }
Here the support is understood with the limiting prescription described below
eqns.~(\ref{eq:lim1}),(\ref{eq:lim2}).
 Hence
\be
Z_{(P,Q)}=
\sum_{r,s\ge 0 }
F_2(r,s;P,Q)+\sum_{ u\geq 0}F_1(u,P+Q)
\label{eq:finalPQ}
\ee
Notice that the second term is simply $Z_{(P+Q)}$.



%
%
%

\subsection{General case}
\label{sec:general}

The evaluation of formula eq.~(\ref{eq:level2}) in the general case can now be carried out
straightforwardly. In particular,  one can construct a $k$ dimensional map $\rho$ from 
$D_\rho\subset \mathbb Z^k$ onto the set of Young diagrams
 by adding further border strips. 
The hook product can again be given in closed form, see appendix (\ref{sec:hookgen}).
%
We find
\be
Z_\lambda=\sum_{s\in D_\rho}
F_k(s;\l)
\ee
where $D_\rho$ is now given in eq.~(\ref{eq:Drhok}), and 
\be
F_k(s;\l)\equiv\prod_{1\leq i\leq k} F_1(s_i,\l_i) \prod_{1\leq i<j\leq k} X(s_i,s_j,\l_i,\l_j)
\label{eq:Fk}
\ee
where $s=(s_1\dots s_k)$ and $\l=(\l_1\dots \l_k)$.

As before, we can replace the set $D_\rho$ by the positive orthant, with some additional lower dimensional terms. To this end, define 
\be
\tilde Z_\l\equiv 
\sum_{s_i\geq 0} F_k(s;\l)
\ee
which is just $Z_\lambda$  but with $D_\rho$ replaced by the entire positive orthant.
$Z_\l$ can be found in terms of $\tilde Z_\l$ of lower or equal dimension as follows.
Define the linear operator $L_{\l_k}$ that adds a  strip of length $\l_k$ to the existing domain. Hence
\be
L_{\l_k}Z_{(\l_1\dots \l_{k-1})}= Z_{(\l_1\dots \l_k)}
\ee
The same operator, when acting on the positive orthant only, produces also some $k-1$ dimensional terms
\footnote{
To obtain again a valid Young diagram in the lower dimensional terms, we should reshuffle the $\l_i$ such that they appear again in decreasing order.
However, since $\tilde Z $ is symmetric in its arguments, the order  does not matter.
}
\be
L_{\l_k}\tilde Z_{(\l_1\dots \l_{k-1})}=\tilde Z_{(\l_1\dots\l_k)}+
\sum_{i=1}^{k-1} \tilde Z_{(\l_1\dots \l_i+\l_k\dots \l_{k-1})}
\label{eq:shifttilde}
\ee
Eq.~(\ref{eq:shifttilde}) follows straightforwardly from the $k$ dimensional analogue of eq.~(\ref{eq:posneg}), eq.~(\ref{eq:rel3}).
We can build any $Z_\l$ by acting repeatedly on the relation 
\be
 Z_{(\l_1)}=\tilde Z_{(\l_1)}
\ee
with the operator $L$.
For instance,
\be
\begin{aligned}
Z_{(\l_1,\l_2,\l_3)}={}& L_{\l_3}L_{\l_2}\tilde Z_{(\l_1)}\\
			={}& L_{\l_3}\left(\tilde Z_{(\l_1,\l_2)}+\tilde Z_{(\l_1+\l_2)}\right)\\
			={}& \tilde Z_{(\l_1,\l_2,\l_3)}+\tilde Z_{(\l_1+\l_3,\l_2)}+ \tilde Z_{(\l_1,\l_2+\l_3)}+\tilde Z_{(\l_1+\l_2,\l_3)}+\tilde Z_{(\l_1+\l_2+\l_3)}
\end{aligned}
\ee
etc.

Formally, the solution of this recursion can be written in terms of partitions on the index set $\{1\dots k\}$.
let \(\Pi_k\) denote the set of  set-partitions of
$\{1,2,\dots k\}$.  For \(\pi=\{B_1,\ldots,B_\ell\}\in\Pi_k\), define
\(
\lambda_B=\sum_{i\in B}\lambda_i
\)
for each block $B\in\pi$, and write
\(
\lambda/\pi=(\lambda_{B_1},\ldots,\lambda_{B_\ell}) .
\)
In words, $\lambda$ mod $\pi$ is the Young diagram obtained from $\l$ by joining  its rows  according to the chosen partition $\pi$. Then the final result is
\be
Z_\lambda
=
\sum_{\pi\in\Pi_k}
\tilde Z_{\lambda/\pi}.
\label{eq:final}
\ee
We have cross-checked eq.~(\ref{eq:final}) against  eq.~(\ref{eq:Zlambda2}) by evaluating the permutation sums in the latter by brute force
for a variety of partitions $\lambda$, finding agreement in all cases.

\section{Discussion}

The final  expression for $Z_\lambda$, derived in section \ref{sec:general} for integer powers $\lambda_i$, can straightforwardly be evaluated at arbitrary real positive $\lambda_i$ by interpreting factorials and Pochhammer symbols as
Gamma functions.\footnote{This is unlike the  expressions eqns.~(\ref{eq:Zlambda}), (\ref{eq:Zlambda2}) 
and (\ref{eq:level2}) which cannot directly be evaluated for non-integer values in any obvious way.}
This does of course not mean that this continuation is unique, as one could for instance
add functions that vanish on the integers. However, we make the following observations:
\begin{itemize}
\item
For $N\in \mathbb R^+$,  $Z_{(N)}/Z_{(1^N)}$ is a completely monotone function of $N$,
as required from the left hand side of eq.~(\ref{eq:singletrace}).
The only conceivable smooth additive corrections that vanish on the integers  
are oscillating functions such as $\sin \pi N$ which spoil this property (the same holds true for 
multiplicative corrections that equal one on the integers).\footnote{Actually, complete monotonicity already allows one to conclude
uniqueness of the continuation. This is a consequence of Bernstein’s theorem plus uniqueness in the Hausdorff moment problem.}
\item
The special case $n=1$ results in a  pure state $\rho_{\text A,\psi}$, and hence  
$ \overline{\tr[\rho_{A}^N]}=1$, for all positive real  $N$. This limiting case is reproduced
by our expressions.
\item
Let $q_s$ be the $m$ eigenvalues of $\rho_{\text A,\psi}$, then $Z_\lambda/Z_{{(1^N)}}$ is the average of 
$\sum_{\{s_i\}} \prod_{i=1}^k q_{s_i}^{\l_i}$,
From eq.~(\ref{eq:final}) alone one can in fact deduce that $\tilde Z_\lambda/Z_{{(1^N)}}$ is the same 
average, except for all eigenvalues different. As such, $\tilde Z_\lambda$ is expected to be zero 
for $k>m$. This is indeed the case for our expressions, since the $k$ indices $s_i$ must all be different for 
all $X$ factors in eq.~(\ref{eq:Fk}) to be nonzero.
\item
We have verified numerically that our result coincides with reference \cite{Bianchi:2019stn} for real 
$\l_i$, up to $k=3$, 
even though the translation is highly nontrivial. We  emphasize that our expressions 
are  simpler than the ones obtained in \cite{Bianchi:2019stn}, for instance, $Z_{(N)}$ is only a single sum rather than a double one, while $Z_{(P,Q)}$ is only a double sum (instead of a quadruple one) etc.
\end{itemize}

Finally, let us comment on some further analytical properties of  $Z_\l$.
\begin{itemize}
\item
As remarked earlier, all sums are cut off at $\min(m,n)-1$ due to the factors 
$\pochb{m-r}{\l_i}\pochb{n-r}{\l_i}$. 
\item
$Z_{(N)}$ can be written as a hypergeometric function at unit argument
\be
Z_{(N)}=\frac{ \poch{m}{N}\poch{n}{N}}{\Gamma(N+1)}
\HGF{1-m,1-n,1-N}{1-m-N,1-n-N}{1}
\ee
Recall that hypergeometric functions are finite sums if any of their upper arguments are negative integers, consistent with the previous point.

\item
$Z_{(N)}/Z_{(1^N)}$ is a rational function of $N$, as it is a finite sum of terms of the form
\be
\sim \frac{\Gamma(N+m-r)\Gamma(N+n-r)}{N\Gamma(N-r)\Gamma(N+mn)}
\ee 
 which are rational since $m,r,n\in \mathbb Z$.
%
%
It can be seen to have poles at the negative integers: let, w.l.o.g., $m\leq n$, then there are precisely
  $mn-n+m-1$ poles at values
$1-mn\leq N
\leq m-n-1$.
\item
The higher power moments $Z_\l/Z_{(1^N)}$ are not rational functions of the $\l_i$, however, 
${Z_\l}/{\Gamma(\l_1)\cdots \Gamma(\l_k)}$ and ${\Gamma(N)}/{Z_{(1^N)}}$ are.
\item
Differentiation of $Z_{(N)}/Z_{(1^N)}$ with respect to $N$ immediately gives the Page curve, eq.~(\ref{eq:page}). 
Differentiation of the generalized power moments gives the higher moments, we have verified the 
existing results in the literature \cite{Page:1993df,Vivo:2016ref,Bianchi:2019stn}. It would be interesting to investigate if our general formula 
can result in closed expressions for all higher moments. However, already the variance is a 
somewhat lengthy calculation with many (somewhat mysterious) cancellations. We leave the derivation of closed form expressions for the higher Page moments to 
future work. 
\end{itemize}

\section{Conclusions}

In this paper we have derived explicit expressions for the power moments eq.~(\ref{eq:trrhoN}) 
of the reduced density matrix of a bipartite pure state, under the assumption that the latter is uniformly distributed over all normalized states.
The derivation is entirely algebraic, leveraging invariance of the measure under unitary transformations.
The key technical tool is Schur-Weyl duality that leads us to the nontrivial rewriting eq.~(\ref{eq:level2}).
The latter can then be explicitly evaluated by means of the Murnaghan-Nakayama rule for the characters of the symmetric group.
Our final expressions, presented in section \ref{sec:general}, coincide numerically with the ones
derived in ref.~\cite{Bianchi:2019stn}, but are somewhat simpler (involving fewer nested sums).
The moments of the entanglement entropy correspond to derivatives of the power moments at the special values $\lambda_i=1$.

\section*{Acknowledgements}
I would like to thank Arman Esmaili for suggestions on the manuscript.
I acknowledge financial support by the Conselho Nacional de Desenvolvimento
Científico e Tecnológico (CNPq) under fellowship number 313238/2023-5, as well as 
the Fundação de Amparo à Pesquisa do Estado do Rio de Janeiro (FAPERJ) under 
project number 210.785/2024.

\appendix

\section{Basics of the symmetric group $S_N$}
\label{sec:SN}


We start by recalling that the symmetric group $S_N$ is defined as the group of permutations of $N$ elements.
Every permutation can be broken up into {\em cycles}, disjoint sets of elements  that cyclically permute amongst themselves.  
For instance
 the permutation $12345\to 41523$ contains a cycle of length three: ($124\to 412$, or $1\to4\to2\to 1$, or in "cycle notation" (142)) and a cycle of length two ($35\to 53$, or simply (35)).
Two group elements $p$ and $p'$  are said to be {\em conjugate} if they can be written as 
$p=qp'q^{-1}$ for some group element $q$. The corresponding equivalence classes are called {\em conjugacy classes}.
For  the group $S_N$, the conjugacy classes
are rather simple to characterize: two elements are conjugate if and only if they are of the same cycle types (the cycle type of the previous example would be $(3,2)$, i.e., one three-cycle and one two-cycle. 
We will always denote them in decreasing order).
Therefore, the conjugacy classes of $S_N$ are in one-to-one correspondence with the 
partitions of $N$. For instance, the group $S_3$ has three equivalence classes
corresponding to the partitions/cycle types $(3)$, $(2,1)$, $(1,1,1)$, having two, three and one 
member(s) respectively. Note that for any $N$, the identity element is the only member of the 
conjugacy class $(1^N)$.

Next, recall that the {\em character} of a group element $p$, in the irreducible representation $\rho$, is defined by $\chi_\rho(p)\equiv \tr \rho(p)$. This is clearly a class function (it is constant on each conjugacy class).
Characters are useful as they furnish a basis for  class functions, that is, any class function $f$ can be written as $f(p)=\sum_\rho c_\rho \chi_\rho(p)$ where $\rho$ runs over all the irreducible representations of $S_N$. The characters of the symmetric group are integers.

In fact, it turns out that the irreducible representations of $S_N$ are also
in one-to-one correspondence with the partitions of $N$. The characters $\chi_\rho(\lambda)$, 
can therefore be interpreted as the components of a square matrix (sometimes denoted as $\chi^\rho_\lambda$). 
In fact, the weighted characters $\chi_\rho(\lambda)\sqrt{|C_\lambda|/{N!}}$ (where $|C_\lambda|$ denotes the
cardinality of the equivalence class $\lambda$) form an orthogonal matrix.

{\em Young diagrams} provide a pictorial way of representing the partitions and irreducible representations.
Given a partition $\lambda=(\lambda_1,\dots,\lambda_k)$ of $N$ with $\lambda_i\geq \lambda_{i+1}$, a Young diagram consists of $N$ cells or boxes, arranged in  $k$ rows  of length $\lambda_i$. For instance, the partitions $\lambda=(4,1,1)$ and $\lambda=(5,3,2)$ are represented by the diagrams
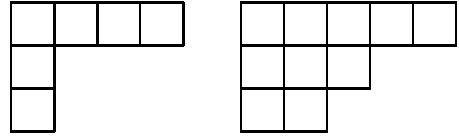
\[
\begin{ytableau}
{}&&&\\
{}\\
{}
\end{ytableau}
\qquad
\begin{ytableau}
{}&&&&\\
&&\\
&
\end{ytableau}
\]
Partitions of the form $(N-r,1^r)$ (like the one on the left) are called {\em hook partitions}.

Let $i$ and $j$ label the row and column of a cell. Two important quantities related to a cell that frequently occur are  its "content"  (defined as $j-i$), and its "hook length"  (defined as the 
the total number of cells contained in its "hook." This includes the cell itself, all cells directly to its right in the same row, and all cells directly below it in the same column).
The contents and the hook lengths of the partition $(5,3,2)$ are thus:
\[
\begin{ytableau}
0&1&2&3&4\\
-1&0&1\\
-2&-1
\end{ytableau}
\qquad 
\begin{ytableau}
7&6&4&2&1\\
4&3&1\\
2&1
\end{ytableau}
\]

\section{Schur-Weyl Duality}
\label{eq:SW}

\label{sec:SW}
We can derive eq.~(\ref{eq:level2}) by making the following observation.
Start from $Z_\lambda$ in the form
\be
Z_\lambda=\sum_{p\in S_N} 
\tr T_n(p)\tr T_m(pq)
\qquad q\in C_\lambda
\label{eq:tensortraces}
\ee
where  the tensor 
\be
[T_n(p)]^{a_1\cdots a_N}_{b_1\cdots b_N}\equiv \prod_{k=1}^N\delta^{a_k}_{b_{p(k)}}
\ee
 can  naturally be interpreted
as the endomorphism that acts on the vector space 
\be
\mathbb C^{n^N}= \mathbb C^n\otimes \mathbb C^n\otimes \dots\otimes\mathbb C^n \qquad \text{($N$ factors)}
\ee
by permuting the $N$ factors according to $p$.
Similarly, 
$T_m(pq)$
acts on $\mathbb C^{m^N}$ 
with the permutation $pq$.

Besides the natural $S_N$ action, $\mathbb C^{n^N}$ also has a natural 
U$(n)$ action which is the diagonal one, i.e., all factors are transformed with the same
unitary transformation. These two actions commute. 
 Schur-Weyl duality  states that $\mathbb C^{n^N}$  can be decomposed into irreducible modules as follows:
\be
\mathbb C^{n^N}=\oplus_\mu (V_{\text{U}(n),\mu}\otimes V_{S_N,\mu})\,,
\label{eq:SWduality}
\ee
where $V_{G,\mu}$ denotes the module of the  irreducible  $G$ representation labeled by $\mu$. 
Recall that  irreducible $S_N$ representations are in one-to-one correspondence with Young diagrams with $N$ boxes, whereas  irreducible U$(n)$ representations are in one-to-one correspondence 
with Young diagrams with at most $n$ rows.
The sum in eq.~(\ref{eq:SWduality})
is therefore restricted to Young diagrams $\mu$ with $N$ boxes and up to $n$ rows.

We can then apply Schur-Weyl duality, eq.~(\ref{eq:SWduality}) to each factor of the summands in  eq.~(\ref{eq:tensortraces}), e.g.
%
\be
\tr_{\mathbb C^{n^N}} T_n(p)=\sum_\mu \tr_{V_{U(n),\mu}}(\mathbf 1)\tr_{V_{S_N,\mu}}(p)
=\sum_\mu
 \dim [V_{U(n),\mu}]\, \chi_\mu(p)
\ee 
and similarly for $\tr_{\mathbb C^{m^N}} T_m(pq)$. The dimension of the U$(n)$ representation can be computed from the hook-content formula \cite{Fulton:1997YoungTableaux}
\be
\dim [V_{U(n),\mu}]=\frac{C_\mu(n)}{H_\mu}
\ee
where 
$C_\mu(n)$ and $H_\mu$ 
were given in eqns.~(\ref{eq:content}) and (\ref{eq:hook}).
Notice that for $\mu=(\mu_1\dots \mu_k)$, 
$C_\mu(n)=0$ for $k> n$.



%
Then
\be
Z_\lambda
	=\sum_{\mu,\nu\vdash N}
	\frac{C_\mu(m)C_\nu(n)}{H_\mu H_{\nu}}
	\sum_{p\in S_N}\chi_\mu(pq)\chi_\nu(p)	
\ee
Finally, we use  character orthogonality (see \cite{isaacs1994character}, theorem 2.13).
\be
\sum_{p\in S_N}\chi_\mu(pq)\chi_\nu(p)=H_\mu \chi_\mu(q) \delta_{\mu\nu}
\ee
to arrive at eq.~(\ref{eq:level2}).

\section{Beta sets and hook length products}

\label{sec:beta}
\subsection{Beta sets}

Beta-sets are a convenient way of constructing valid Young diagrams $\mu$ by successively adding border strips
to smaller Young diagrams \cite{james1981representation}. Representing Young diagrams this way makes it very easy to evaluate the three quantities $\chi_\mu(\lambda)$, $H_\mu$, and $C_\mu$ appearing in our main formula eq.~(\ref{eq:level2}).
We illustrate this by  constructing a partition of size 
$N=P+Q$ out of the empty partition by first adding a border strip of length $P$ and then one of length $Q$, before giving the general construction.
This will give explicitly the map $(r,s)\mapsto \rho(r,s)$, including its somewhat nontrivial domain $D_\rho$.

We use finite beta-sets of length \(N\).  For a partition $\lambda$ of $N$, define a set of $N$ integers as follows.\footnote{Notice that we use beta sets that are shifted by $N-1$ with respect to the standard convention.}
\be
B_\lambda\equiv \{\beta_i \ |\ 1\leq i\leq N\},
\label{eq:defbeta}\qquad \text{with\ }\beta_i\equiv \lambda_i-i+1
\ee
where the partition is padded with zeros to make a Young diagram with exactly $N$ rows. 
Notice that the $\beta_i$ form a strictly decreasing sequence $\beta_1>\beta_2>\dots$.
The mapping $\lambda\mapsto B_\lambda$ is bijective, i.e., from a valid ordered beta set one can always obtain the partition by $\lambda_i=\beta_i+i-1$.

We now want to construct the partition/beta-set corresponding to adding a border strip 
of length $P$
with LLC content $-r$ to the empty partition.
Start from the "vacuum" beta-set
\be
B_\varnothing=\{0,-1,-2,\dots,-N+1\}.
\ee
which according to  eq.~(\ref{eq:defbeta}) is indeed the partition with all $\lambda_i=0$. 
A bead move is an operation that substitutes $a\in B_\lambda$ with $a'\notin B_\lambda$.
It corresponds to adding a valid 
border strip to the partition 
(the new beta set corresponds to a valid partition/Young diagram).
For the case at hand,
a border strip of length $P$ obtained by the bead move 
\be
a=-r
\ \to \
a'=-r+P.
\ee
For it to be legal (i.e., $a\in B_\varnothing$, $a'\notin B_\varnothing)$, we need 
\be
r\in D_1\equiv \{0,1,\dots P-1\}\,,
\ee
 precisely 
resulting  in all the Young diagrams of hook shape of height $r$, 
with LLC  content $-r$. 

Next, a border strip of length $Q$ whose LLC cell has content $-s$ is obtained by the bead move
\be
b=-s
\ \to\ 
b'=-s+Q.
\ee
Its legality is ensured if $-s\in B_{\rho(r)}$ and $-s+Q\notin B_{\rho(r)}$, in other words
\be
s\in D_2\equiv  \{0,1\dots Q-1\}\cup \{r-P,r+Q\} \backslash \{r,r+Q-P\}
\label{eq:S2}
\ee 
The  extra elements in eq.~(\ref{eq:S2}) correspond to $b=a'$ and $b'=a$, while the removed elements correspond to $b=a$, $b'=a'$.
The two successive moves $a\to a'$ and $b\to b'$ define the map $\rho(r,s)$ along its domain $D_\rho$
\be
D_\rho=\{(r,s)\ | \ r\in D_1, s\in D_2\}\,.
\label{eq:Drho2}
\ee

Let us illustrate it by some examples with $P=3$, $Q=2$.
First consider the sequence of moves
\be
0\ \to \ 3\,,\qquad -1\ \to +1
\ee
that is, $r=0$, $s=1$. This gives:
\[
\renewcommand{\arraystretch}{1.5}
\begin{array}{c@{\qquad}c@{\qquad}c}
B_{\varnothing}
&
B_{(3)}
&
B_{(3,2)}
\\[4pt]
\{0,-1,-2,-3,-4\}
&
\{+3,-1,-2,-3,-4\}
&
\{+3,+1,-2,-3,-4\}
\\[8pt]
\begin{ytableau}
*(gray!20)0\\
*(gray!20)-1\\
*(gray!20)-2\\
*(gray!20)-3\\
*(gray!20)-4
\end{ytableau}
&
\begin{ytableau}
\colA0&\colA1&\colA2&*(gray!20)3\\
*(gray!20)-1\\
*(gray!20)-2\\
*(gray!20)-3\\
*(gray!20)-4
\end{ytableau}
&
\begin{ytableau}
\colA0&\colA1&\colA2&*(gray!20)3\\
\colB-1&\colB0&*(gray!20)1\\
*(gray!20)-2\\
*(gray!20)-3\\
*(gray!20)-4
\end{ytableau}
\end{array}
\]
Here, we also displayed  "ghost cells" in gray, which are {\em not} part of the Young diagram, but rather  prospective cells for the LLCs of possible border
strips, and which precisely contain the values of the beta set. 

As a second example, consider the moves
\be
-2\ \to\ +1\,,\qquad +1\ \to +3
\label{eq:hookex2}
\ee
in other words, $r=2$ and $s=-1$.
\[
\renewcommand{\arraystretch}{1.5}
\begin{array}{c@{\qquad}c@{\qquad}c}
B_{\varnothing}
&
B_{(1,1,1)}
&
B_{(3,1,1)}
\\[4pt]
\{0,-1,-2,-3,-4\}
&
\{+1,0,-1,-3,-4\}
&
\{+3,0,-1,-3,-4\}
\\[8pt]
\begin{ytableau}
*(gray!20)0\\
*(gray!20)-1\\
*(gray!20)-2\\
*(gray!20)-3\\
*(gray!20)-4
\end{ytableau}
&
\begin{ytableau}
\colA0 & *(gray!20)1\\
\colA-1 & *(gray!20)0\\
\colA-2 & *(gray!20)-1\\
*(gray!20)-3\\
*(gray!20)-4
\end{ytableau}
&
\begin{ytableau}
\colA0 & \colB1 & \colB2 & *(gray!20)3\\
\colA-1 & *(gray!20)0\\
\colA-2 & *(gray!20)-1\\
*(gray!20)-3\\
*(gray!20)-4
\end{ytableau}
\end{array}
\]
Notice that the second move is starting precisely at the end point of the first.

A third example is
\be
-1\ \to\ +2\,,\qquad -3\ \to -1
\label{eq:hookex3}
\ee
or $r=1$, $s=3$,
giving
\[
\renewcommand{\arraystretch}{1.5}
\begin{array}{c@{\qquad}c@{\qquad}c}
B_{\varnothing}
&
B_{(2,1)}
&
B_{(2,1,1,1)}
\\[4pt]
\{0,-1,-2,-3,-4\}
&
\{+2,0,-2,-3,-4\}
&
\{+2,0,-1,-2,-4\}
\\[8pt]
\begin{ytableau}
*(gray!20)0\\
*(gray!20)-1\\
*(gray!20)-2\\
*(gray!20)-3\\
*(gray!20)-4
\end{ytableau}
&
\begin{ytableau}
\colA0 & \colA1 & *(gray!20)2\\
\colA-1 & *(gray!20)0\\
*(gray!20)-2\\
*(gray!20)-3\\
*(gray!20)-4
\end{ytableau}
&
\begin{ytableau}
\colA0 & \colA1 & *(gray!20)2\\
\colA-1 & *(gray!20)0\\
\colB-2 & *(gray!20)-1\\
\colB-3 & *(gray!20)-2\\
*(gray!20)-4
\end{ytableau}
\end{array}
\]
This time, the second move ends at the starting point of the first.

\subsection{Calculating the hook length product from beta sets}

The hook product can be conveniently computed from the beta set elements as follows \cite{james1981representation,macdonald1998symmetric}:
\be
\frac{1}{H_\lambda}
=
\frac{
\prod_{1\le i<j\le N}(\beta_i-\beta_j)
}{
\prod_{i=1}^N (\beta_i+N-1)!
}.
\label{eq:hookprodbeta}
\ee
One may verify immediately from eq.~(\ref{eq:hookprodbeta}) that the hook product of the empty partition is $H_\varnothing=1$, consistent with the fact
that it should be the empty product (Young diagram with no boxes).
Therefore, 
 the inverse hook product of \(\lambda\) is determined by the change in the denominator  and the change in the  numerator (the "Vandermonde" factor").
The former part gives
\be
\frac{(a+N-1)!}{(a'+N-1)!}=\frac{1}{(N-r)^{\overline P}}.
\ee
The Vandermonde part of the $a\to a'$ move contributes
\be
\prod_{ \substack{x\in B_\varnothing\\x\neq a}}^{N-1}\left|\frac{a'-x}{a-x}\right|=\frac{(P+1)^{\overline {N-r-1}}(P-r)^{\overline {r}}}{(N-r-1)!r!}
\ee
Combining the factorial and the Vandermonde terms we get  the contribution
\be
\frac{(N-r-1)!}{(N-r+P-1)!}\frac{(P+1)^{\overline {N-r-1}}(P-r)^{\overline {r}}}{(N-r-1)!r!}
=\frac{1}{P(P-r-1)!r!}
\label{eq:hook1}
\ee
It is reassuring that this reproduces the correct hook-length product of a pure hook shape.
An independent $b\to b'$ move would contribute
\be
\frac{1}{Q(Q-s-1)!s!}
\ee
However, $b\to b'$ does not start from the vacuum set, but rather from $B_{\rho(r)}$.
We can simply correct for this by modifying the
Vandermonde terms by an additional factor:
\be
\frac{\frac{a'-b'}{a-b}}{\frac{a'-b}{a-b}\frac{a-b'}{a-b}}=
\frac{(a'-b')(a-b)}{(a'-b)(a-b')}
=
\frac{
(Q-P+r-s)(s-r)
}{
(Q-s+r)(P+s-r)
}
\equiv X(r,s,P,Q)
\label{eq:interaction}
\ee
where the numerator on the left hand side contains the true effect of $(a,b)\to(a',b')$ in the
Vandermonde factor, and the denominator removes again the ratios that were "erroneously" included because the second move  was considered independent from the first.

Combining the two independent bead contributions with the interaction factor gives
\be
\frac{1}{H_{\rho(r,s)}}
=
\left|
\frac{1}{P (P-r-1)!r!}
\frac{1}{Q (Q-s-1)!s!}
\frac{
(s-r)(Q-P+r-s)
}{
(Q-s+r)(P+s-r)
}\right|.
\label{eq:hookprodfinal}
\ee
This is the desired hook-product formula in the $r,s$ coordinates.
The absolute value is needed since the cross-ratio is not always positive and the hook 
product by definition is positive (in the considerations above we have not paid attention to signs).
We will come back to the question of signs in a moment.

In the  special cases  $b=a'$ and $b'=a$ the two moves $a\to a'$ and $b\to b'$ 
combine to a single effective move.
To recover the expected results from eq.~(\ref{eq:hookprodfinal}) we must interpret the 
factorials as Gamma functions, such that its poles cancel the poles from the interaction terms.
Indeed, for $b=a'$ the corresponding limit  $s\to r-P$ in eq.~(\ref{eq:hookprodfinal}) gives
\be
\frac{1}{H_{\rho(r,s)}}\to \frac{1}{N(N-r-1)!r!}
\label{eq:lim1}
\ee
which is precisely the result expected from the hook shape obtained by the 
single move
$
a=-r\ \to\ b'=-r+N$, 
see the example eq.~(\ref{eq:hookex2}) above.
Similarly, when $b'=a$,  the limit $s\to r+Q$ in eq.~(\ref{eq:hookprodfinal}) gives
\be
\frac{1}{H_{\rho(r,s)}} \to \frac{1}{N(N-s-1)!s!}
\label{eq:lim2}
\ee
which is again the expected result, compare with the example eq.~(\ref{eq:hookex3}) above.

In the following, whenever the closed expressions for $H^{-1}$ are evaluated on such
special points (i.e. where bead moves collide and generate spurious poles), they are understood in this limiting sense.

Finally, we must figure out the sign. 
%
We claim that    
\be
(-1)^{\h (r)+\h (s)}\frac{1}{H_{\rho(r,s)}}
=(-1)^{r+s}
\frac{1}{P (P-r-1)!r!}
\frac{1}{Q (Q-s-1)!s!}
X(r,s,P,Q)
\label{eq:hookprodsigned}
\ee 
The  reason for this is that the height of each strip is given by the number of bead crossings, i.e., the number of beads the move $a\to a'$ (or $b\to b'$) jump in the beta set. When starting from the vacuum beta set
we jump exactly $r$ ($s$) beads. 
The number of bead crossings for the second move is smaller (larger) by one if the $b\to b'$ move jumps over the bead $a$ ($a'$). This sign change is precisely accounted for by the 
sign of the cross ratio, resulting in eq.~(\ref{eq:hookprodsigned}).

The summand in eq.~(\ref{eq:rewriting}) is thus given explicitly as
\be
F_2(r,s;P,Q)\equiv F_1(r,P)F_1(s,Q)X(r,s,P,Q)
\ee
It has the obvious symmetry property
\be
F_2(r,s;P,Q)=F_2(s,r;Q,P)
\ee
Moreover, due to the limits eqns.~(\ref{eq:lim1}) and (\ref{eq:lim2}), we have, for $r\ge 0$:
\be
\begin{aligned}
F_2(r,r-P;P,Q)={}&F_1(r,P+Q)\label{eq:rel1}\\
F_2(r,r+Q;P,Q)={}&-F_1(r+Q,P+Q)
\end{aligned}
\ee

\subsection{Generalization to $k\geq 2$ bead moves}
\label{sec:hookgen}

The generalization to $k$ bead moves is straightforward. We just keep adding strips by performing additional bead moves starting at cells with content $-s_i$.
This defines a map $\rho(s)$, where $s=(s_1,\dots,s_k)$ with the  domain
\be
D_\rho=\{  s \ | \ s_i\in D_i \}
\label{eq:Drhok}
\ee
where, recursively 
\be
D_i\equiv \{0,\dots \lambda_i-1\}\cup \bigcup_{j<i} \{s_j-\l_j, s_j+\lambda_i\}
\setminus \bigcup_{j<i} \{ s_j, s_j-\lambda_j+\lambda_i\} 
\ee
Notice that $D_i$ depends on all the $s_j$ with $j<i$.
The hook product now produces additional cross ratio factors $X$.
We directly give the result for the final summand in the expression $Z_\l=\sum_{s\in D_\rho} F_k(s;\l)$,
\be
F_k(s;\l)
=\prod_{i\leq k} F_1(s_i,\l_i)\prod_{1\leq i<j\leq k}X(s_i,s_j,\l_i,\l_j) 
\ee 
Notice the manifest symmetry of $F_k$ under the simultaneous exchange
\be
s_i\leftrightarrow s_j\qquad \l_i\leftrightarrow \l_j
\ee
Besides eq.~(\ref{eq:rel1}) we have
\be
\begin{aligned}
X(s_i,s_\ell,\l_i,\l_\ell)X(s_i,s_\ell-\l_\ell,\l_i,\l_k)={}&X(s_i,s_\ell,\l_i,\l_\ell+\l_k)\\
X(s_i,s_\ell,\l_i,\l_\ell)X(s_i,s_\ell+\l_k,\l_i,\l_k)={}&X(s_i,s_\ell+\l_k,\l_i,\l_\ell+\l_k)
\label{eq:rel2}
\end{aligned}
\ee
Eqns.~(\ref{eq:rel1}) and (\ref{eq:rel2}) then imply the relation
\begin{multline}
\sum_{\substack{0\le s_\ell<\l_\ell\\ s_k=s_\ell-\l_\ell}} F_k(\dots s_\ell \dots s_k\, ; \dots\l_\ell \dots\l_k)=
\sum_{\substack{0\leq s_k< \l_k\\ s_\ell=s_k+\l_\ell}}F_k(\dots s_\ell \dots s_k\, ; \dots\l_\ell \dots\l_k)\\
+\sum_{0\le s_\ell<\l_\ell+\l_k} F_{k-1}(\dots s_\ell\dots ; \dots\l_\ell+\l_k\dots)
\label{eq:rel3}
\end{multline}
which is the higher dimensional analogue of eq.~(\ref{eq:posneg}).

%


\bibliography{literature} 
 
\end{document}